\documentclass[12pt,a4paper]{article}
\usepackage{amsmath}
\usepackage{amsfonts}
\usepackage{amssymb}
\usepackage{graphicx}

\begin{document}

\author{Vladimir K. Petrov\thanks{ E-mail address: vkpetrov@yahoo.com}}
\title{Fermion propagator at finite temperatures on extremely anisotropic lattice.}
\date{\textit{\ Bogolyubov Intitute for Theoretical Physics, 252143 Kiev, Ukraine. }}
\maketitle
\begin{abstract}
Fermion propagator is computed in a simple model on an extremely anisotropic
lattice $\xi\gg1$. Fermion determinant is evaluated up to $\xi^{-4}$ order.
Chiral condensate is estimated in mean field approximation.
\end{abstract}

\section{ Introduction}

Finite-temperature studies of the fermion contribution in lattice gauge
theories were on the agenda over a number of years and both MC experiments and
analytical models were used to challenge the problem. Calculations with
dynamical fermions and small lattice spacing, however, are still highly
expensive. Hence the approximations which would hopefully capture some of the
essential features of the physics are worth a try. We attempt to study the
fermionic part of the action (see e.g.\cite{HK})
\begin{equation}
-S_{F}=n_{f}a^{3}\sum_{x,x^{\prime}}\left(  \overline{\psi}%
_{x^{\prime}}D_{x^{\prime}x}^{0}\psi_{x}+\xi^{-1}\!\!\!\ \overline{\psi
}_{x^{\prime}}\sum_{n=1}^{3}D_{x^{\prime}x}^{n}\psi_{x}\right) \label{sf}%
\end{equation}
on an extremely anisotropic lattice ($\xi\gg1$) in a simple model \cite{P},
where terms in $\left(  \ref{sf}\right)  $ proportional to $\xi^{-1}$\ are
discarded. Here $n_{f}$\ is the number of flavors and $\xi=$ $\xi\left(
g,\xi_{0}\right)  $ is the anisotropy parameter. The dependance of $\xi$ on
coupling $g$ and on 'bare' anisotropy parameter $\xi_{0}=a/a_{\tau}\ $is
defined by the condition of independence of physical values on
spatial\footnote{Further for convenience we put $a=1$.} $a$ and temporal
$a_{\tau}$ lattice spacings. Dirac operator is defined by
\begin{equation}
D_{x^{\prime}x}^{\nu}=\frac{1-\gamma_{\nu}}{2} U_{\nu}\left(
x\right)  \delta_{x,x^{\prime}-\nu}+\frac{1+\gamma_{\nu}}{2}%
U_{\nu}^{\dagger}\left(  x^{\prime}\right)  \delta
_{x,x^{\prime}+\nu}--\left(  1+\delta_{\nu}^{0}ma_{\tau}\right)
\delta_{x^{\prime},x},\label{D}%
\end{equation}
where $U_{\nu}\left(  x\right)  $ are link $\left(  x,x+\nu\right)  $
variables, $\gamma_{\nu}$ are Dirac matrices, and $\nu=0,1,2,3$. We consider
the case of infinitely heavy mirror fermions and choose Hamiltonian gauge
\begin{equation}
U_{0}\left(  \mathbf{x},t\right)  _{\alpha\lambda}=\delta_{\alpha\lambda
}\left[  1-\left(  \exp\left\{  -i\varphi_{\alpha}\left(  \mathbf{x}\right)
\right\}  +1\right)  \delta_{t^{\prime},0}\right]  ,
\end{equation}
where $\exp\left\{  -i\varphi_{\alpha}\left(  \mathbf{x}\right)  \right\}
\equiv\left(  \prod_{t=0}^{N_{\tau}-1}U_{0}\left(  \mathbf{x},t\right)
\right)  _{\alpha\alpha}$ are the eigenvalues of Polyakov matrix and $N_{\tau
}$ is the lattice length in the temporal direction and $\alpha,\lambda
=1,...,N$ are color indices. In particular, for the temporal component of
Dirac matrix one may write
\begin{equation}
D_{x^{\prime}x}^{0}=\delta_{\mathbf{x}^{\prime},\mathbf{x}}\left(
\frac{1-\gamma_{0}}{2}\Delta_{t^{\prime}t}+\frac{1+\gamma_{0}}{2}%
\Delta_{t^{\prime}t}^{\dagger}\right)  ,\label{Do}%
\end{equation}
with
\begin{equation}
\Delta_{t^{\prime}t}\left(  \mathbf{x}\right)  \equiv\exp\left\{
-i\varphi_{\alpha}\left(  \mathbf{x}\right)  \right\}  \delta_{t^{\prime}%
,t-1}-\left(  1+ma_{\tau}\right)  \delta_{t^{\prime},t}.
\end{equation}
For antiperiodic border conditions on fermion fields in temporal direction the
matrix $\Delta_{t^{\prime}t}$ may be diagonalized by discrete Fourier
transformation with half-integer variables, so
\begin{equation}
\Phi_{t}^{\left(  k\right)  }=\exp\left\{  2\pi i\left(  k+1/2\right)  \left(
t+1/2\right)  /N_{\tau}\right\}  \
\end{equation}
are the eigenfunctions of the operator $\Delta_{t^{\prime}t}$. After Fourier
transformation we get
\begin{equation}
\widetilde{\Delta}_{k,k^{\prime}}=\sum_{t,t^{\prime}=0}^{N_{\tau}-1}%
\Phi_{t^{\prime}}^{\left(  k^{\prime}\right)  }\Delta_{t^{\prime},t}\Phi
_{t}^{\left(  k\right)  \ast}=\delta_{k,k^{\prime}}\lambda_{k},
\end{equation}
where
\begin{equation}
\lambda_{k}=\exp\left\{  -i\left[  \varphi_{\alpha}-2\pi\left(  k+1/2\right)
\right]  /N_{\tau}\right\}  -\left(  1+ma_{\tau}\right)
\end{equation}
are the eigenvalues of $\Delta_{t^{\prime}t}$. Thereby the inverse matrix is
given by
\begin{equation}
\Delta_{t^{\prime},t}^{-1}=N_{\tau}^{-1}\sum_{k=0}^{N_{\tau}-1}\lambda
_{k}^{-1}\exp\left\{  2\pi i\left(  t-t^{\prime}\right)  \left(  k+1/2\right)
/N_{\tau}\right\}  .\label{ft}%
\end{equation}
Common properties of projectors $\frac{1\pm\gamma_{0}}{2}=\left(  \frac
{1\pm\gamma_{0}}{2}\right)  ^{2}$ and $\frac{1\mp\gamma_{0}}{2}\frac
{1\pm\gamma_{0}}{2}=0$ allow to write
\begin{equation}
f\left(  \frac{1+\gamma_{0}}{2}u+\frac{1-\gamma_{0}}{2}v\right)
=\frac{1+\gamma_{0}}{2}f\left(  u\right)  +\frac{1-\gamma_{0}}{2}f\left(
v\right)
\end{equation}
for the arbitrary regular function, in particular the \textit{inverse}
operator $\left(  D_{t^{\prime}t}^{0}\right)  ^{-1}$ may be written as
\begin{equation}
\left(  D_{t^{\prime}t}^{0}\right)  ^{-1}=\frac{1+\gamma_{0}}{2}%
\Delta_{t^{\prime}t}^{-1}+\frac{1-\gamma_{0}}{2}\left(  \Delta_{t^{\prime}%
t}^{-1}\right)  ^{\dagger}.\label{Deto}%
\end{equation}

\section{Ansatz}

To sum over the Fourier variables $k$ in $\left(  \ref{ft}\right)  $ we use a
standard trick. Let $f\left(  \omega\right)  $ be a rational function of
$\cos\omega$ and $\sin\omega$, which has no poles on lines $\operatorname{Re}%
\omega=2\pi n$. The integral over the whole area of the complex variable
$\omega$ for the periodic function splits into a set of equal integrals over
'elementary' bands $2n\pi\leq\operatorname{Re}\omega-\frac{\pi}{N_{\tau}}%
<2\pi\left(  n+1\right)  $. Due to periodicity, the result of integration over
any band boundary is
\begin{equation}
\Phi=\frac{1}{2}\left\{  \int_{-i\infty}^{2\pi-i\infty}-\int_{i\infty}%
^{2\pi+i\infty}\right\}  f\left(  \omega\right)  \tan\frac{N_{\tau}\omega}%
{2}\frac{d\omega}{2\pi i}.
\end{equation}
The loop integral over each band area of $\omega$ is equal to zero as well, so
expressing it as sum of residues one may write
\begin{equation}%
{\displaystyle\oint}
f\left(  \omega\right)  \tan\frac{N_{\tau}\omega}{2}\frac{d\omega}{2\pi
i}=\sum res\left\{  f\left(  \omega\right)  \tan\frac{N_{\tau}\omega}%
{2}\right\}  +\Phi=0.
\end{equation}
Taking into account that $\tan\left(  N_{\tau}\omega/2\right)  $ has poles at
$\omega=2\pi\left(  k+12\right)  /N_{\tau}$, one may finally write
\begin{equation}
\frac{1}{N_{\tau}}\sum_{k=0}^{N_{\tau}-1}f\left(  2\pi\left(  k+\frac{1}%
{2}\right)  /N_{\tau}\right)  =-\frac{1}{2}\sum_{\left\{  \omega_{r}\right\}
}res\left\{  f\left(  \omega\right)  \tan\frac{N_{\tau}\omega}{2}\right\}
-\Phi,\label{int}%
\end{equation}
where $\omega_{r}$ poles of the $f\left(  \omega\right)  $ are located in the
band $0\leq\operatorname{Re}\omega-\frac{\pi}{N_{\tau}}<2\pi.$ Now one may
apply the ansatz $\left(  \ref{int}\right)  $ to compute $\Delta_{t,t^{\prime
}}^{-1}$. In the case considered
\begin{equation}
f\left(  \omega\right)  =\left\{
\begin{array}
[c]{ccc}%
-e^{i\left(  t-t^{\prime}\right)  \omega}\lambda_{k}^{-1} & \ for & 0\leq
t-t^{\prime}<N_{\tau}\\
e^{i\left(  t-t^{\prime}+N_{\tau}\right)  \omega}\lambda_{k}^{-1} & for &
-N_{\tau}<t-t^{\prime}<0
\end{array}
\right.
\end{equation}
has poles at $\omega_{0}=-\varphi_{\alpha}+ima_{\tau}$, so taking into account
that $f\left(  \omega\right)  \rightarrow0$ for $\operatorname{Im}%
\omega\rightarrow\pm\infty$ we find
\begin{equation}
\left(  \Delta_{t,t^{\prime}}^{-1}\right)  _{\alpha}=\frac{\theta
_{t,t^{\prime}}e^{-i\varphi_{\alpha}}-\left(  \theta_{t^{\prime},t}%
+\delta_{t^{\prime},t}\right)  e^{\frac{m}{T}}}{e^{-i\varphi_{\alpha}%
}+e^{\frac{m}{T}}}e^{\frac{m}{T}\frac{t-t^{\prime}-1}{N_{\tau}}},\label{prop}%
\end{equation}
where $T=N_{\tau}^{-1}a_{\tau}^{-1}$ and the periodic in $t_{k}$ function
$\theta_{t_{1},t_{2}}$ is given by (for $t_{k}=\left(  t_{k}\right)
_{\operatorname{mod}N_{\tau}}$)
\begin{equation}
\theta_{t_{1},t_{2}}=\left\{
\begin{array}
[c]{ccc}%
1 & for & t_{1}>t_{2}\\
0 & for & t_{1}\leq t_{2}%
\end{array}
\right.  .\quad
\end{equation}
Hence, in our approximation the expression for fermion determinant may be
found in a closed form from $\left(  \ref{Deto}\right)  $ and $\left(
\ref{prop}\right)  $.

\section{ Fermion determinant}

Since $\left(  D^{0}\right)  ^{-1}$is known, for the fermion contribution up
to the $O\left(  \xi^{-4}\right)  $ one can write
\begin{equation}
n_{f}^{-1}\ln\int d\overline{\psi}_{x^{\prime}}d\psi_{x}\exp\left\{
-S_{F}\left(  1/\xi\right)  \right\}  =-S_{F}^{\left(  eff\right)  }\left(
1/\xi\right)  =-S_{F}^{\left(  eff\right)  }\left(  0\right)  -\frac{1}%
{2}S_{F}^{\left(  eff\right)  \prime\prime}\left(  0\right)  \xi^{-2},
\end{equation}
where
\begin{equation}
-S_{F}^{\left(  eff\right)  }\left(  0\right)  =\ln\det\left(  D^{0}\right)
=\sum_{\mathbf{x}}\left\{  \sum_{\alpha=1}^{N}\ln\left(  \cos\varphi_{\alpha
}\left(  \mathbf{x}\right)  +\cosh\frac{m}{T}\right)  -S_{mirr}\right\}
\end{equation}
with
\begin{equation}
-S_{mirr}=Nm/T\label{mir2}%
\end{equation}
and
\begin{equation}
S_{F}^{\left(  eff\right)  \prime\prime}\left(  0\right)  =\sum_{n=1}%
^{3}\mathrm{Sp}\left\{  \left(  D^{0}\right)  ^{-1}D^{n}\left(  D^{0}\right)
^{-1}D^{n}\right\}  .
\end{equation}
In case of light $\left(  m\ll T\right)  $ fermions we get
\begin{align}
-\frac{1}{2}S_{F}^{\left(  eff\right)  \prime\prime}\left(  0\right)   &
=\left(  1-\frac{2m}{T}+O\left(  \frac{m^{2}}{T^{2}}\right)  \right)
\sum_{t_{2,}t_{1}=0}^{N_{\tau}-1}\sum_{\mathbf{x}}\sum_{n=1}^{3}%
\times\nonumber\\
&  \operatorname{Re}\left\{  U_{n}\left(  t_{2},\mathbf{x}\right)  _{\alpha
\nu}\Phi_{\alpha\nu}\left(  \mathbf{x,n}\right)  _{t_{2},t_{1}}U_{n}^{\dagger
}\left(  t_{1},\mathbf{x}\right)  _{\nu\alpha}-\Phi_{\alpha\alpha}\left(
\mathbf{x,n}\right)  _{t_{2},t_{1}}\right\}  .
\end{align}
where%

\begin{align}
\Phi_{\alpha\nu}\left(  \mathbf{x,n}\right)  _{t_{2},t_{1}} &  =-\frac{1}%
{2}\left(  \dfrac{\theta_{t_{1},t_{2}}}{1-e^{-i\varphi_{\nu}\left(
\mathbf{x+n}\right)  }}+\dfrac{\theta_{t_{2},t_{1}}}{1-e^{-i\varphi_{\alpha
}\left(  \mathbf{x}\right)  }}\right)  \times\nonumber\\
&  \tan\dfrac{\varphi_{\alpha}\left(  \mathbf{x}\right)  }{2}\tan
\dfrac{\varphi_{\nu}\left(  \mathbf{x+n}\right)  }{2}%
\end{align}

It is easy to see that $S_{F}^{\left(  eff\right)  \prime\prime}\left(
0\right)  $ turns into zero in the case of $U(1)$ gauge group, if spatial link
variables $U_{n}$ are \textit{static}.

\section{Chiral condensate}

To estimate $\left\langle \overline{\psi}\psi\right\rangle $ one may, in the
spirit of Curie-Weiss method, change all $\varphi_{\alpha}\left(
\mathbf{x}\right)  $ for the mean field $\overline{\varphi}_{\alpha}$ defined
by the mean field equation. For example, in case of SU(2) gauge group
$\varphi_{1}\left(  \mathbf{x}\right)  =-\varphi_{2}\left(  \mathbf{x}\right)
\equiv\varphi\left(  \mathbf{x}\right)  /2\simeq\overline{\varphi}/2$. Thereby
for the chiral condensate we get%

\begin{equation}
\left\langle \overline{\psi}\psi\right\rangle =T\frac{\partial}{\partial m}\ln
Z=n_{f}T\left(  \frac{\partial}{\partial m}\ln\det D\right)  _{\varphi
_{\alpha}\left(  \mathbf{x}\right)  =\overline{\varphi}_{\alpha}}%
\end{equation}
or$\allowbreak$%
\begin{equation}
\left\langle \overline{\psi}\psi\right\rangle /2n_{f}=1+\frac{\xi^{-2}}%
{2}\left.  S_{F}^{\left(  eff\right)  \prime\prime}\left(  0\right)  \right|
_{m=0}+\frac{m/\allowbreak T}{1+\cos\left(  \overline{\varphi}/2\right)
}+O\left(  m^{2},\xi^{-4}\right)  .\label{psi}%
\end{equation}
It should be noted that even for infinitely heavy mirror fermion the remnant
term $S_{mirr}$, defined in $\left(  \ref{mir2}\right)  $survives. It is easy
to make sure that this very term introduces a constant component $\left(
2n_{f}\right)  $ into $\langle\overline{\psi}\psi\rangle$, that doesn't
disappear even at $m/T\rightarrow0$.\ The mirror fermion contribution may be
totally removed ''by brute force''. If we introduce ''the counterterm''
$S_{ct}=-S_{mirr}$ into the original action $\left(  \ref{sf}\right)  $, but
preserve a standard definition for $\langle\overline{\psi}\psi\rangle$\ , such
''counterterm'' would evidently cancel the undesirable component in the first
term of $\left(  \ref{psi}\right)  $. Therefore, we get $\left\langle
\overline{\psi}\psi\right\rangle \rightarrow n_{f}\xi^{-2}\left.
S_{F}^{\left(  eff\right)  \prime\prime}\left(  0\right)  \right|  _{m=0}$for
$m/T\rightarrow0$.

\section{Discussion}

In the parameter area available for modern computers sea quark contribution
introduces minor changes in MC data. For example, the comparison of
spectroscopy results obtained with dynamical and quenched fermions shows no
dramatic difference. Apparently at the parameter area of the MC simulation the
sea quarks simply do not affect the spectroscopy above five to ten per cent
level \cite{B-W}. Major part of MC data\cite{C96-S98} helps to trace the sea
quark contribution, rather than locate the area where the sea quarks should
obligatory be taken into account.

Suggested model allows to evaluate the fermion contribution analytically for
arbitrary small values of lattice spacings and infinitely large lattices,
where MC simulations easily become prohibitively costly. Although such
contribution can be totally incorporated into the invariant measure and may
appear almost trivial, it significantly changes the phase structure of the
model. Such changes, especially in the case of heavy quarks, are very similar
to those introduced by external magnetic field into spin systems. We hope that
suggested model will help not only to estimate the sea quark contribution at
lattices with size and spacings unaccessible for MC experiment, but might also
assist in qualifying the parameter area where the presence of dynamical
fermions leads to particularly appreciable effects.

\end{document}